# A Continuous Opinion Dynamic Model in Co-evolving Networks—A Novel Group Decision Approach

Qingxing Dong, Xin Zhou

## Abstract

Opinion polarization is a ubiquitous phenomenon in opinion dynamics. In contrast to the traditional consensus oriented group decision making (GDM) framework, this paper proposes a framework with the co-evolution of both opinions and relationship networks to improve the potential consensus level of a group and help the group reach a stable state. Taking the bound of confidence and the degree of individual's persistence into consideration, the evolution of the opinion is driven by the relationship among the group. Meanwhile, the antagonism or cooperation of individuals presented by the network topology also evolve according to the dynamic opinion distances. Opinions are convergent and the stable state will be reached in this co-evolution mechanism. We further explored this framework through simulation experiments. The simulation results verify the influence of the level of persistence on the time cost and indicate the influence of group size, the initial topology of networks and the bound of confidence on the number of opinion clusters.

# 1 Introduction

With the increasing complexity of the crucial decisions in business, management and politics, people usually rely on the wisdom of the crowd rather than that of a single intelligent individual. Thus, group decision making (GDM) which combines the intelligence of individuals (experts or agents), is widely used to tackle real world problems [1, 2]. To organize successful GDM, the diversity of opinions and expertise of knowledge, should be addressed [3]. But at the end of a GDM process, the diversity might be reduced and a certain level consensus in the group can be achieved.

The group consensus problem has been widely studied in decision science research. Generally, to reach a sufficient level of agreement and improve the acceptance of the final solution, GDM problems involve a consensus process and selection process [4]. The consensus process which aims to reach a considerable degree of consensus among the individuals has always been achieved in the form of feedback through iterative group negotiations and is implemented automatically or guided by a moderator. Researchers have proposed many approaches to model the consensus process for different representations of preference formats such as utility values [5], preference relations [6], fuzzy preference relations [7], linguistic preference relations [8-10] and pair-wise comparison matrices of the AHP/ANP[11]. From previous models, we can see that the feedback mechanisms are playing a significant role in improving the consensus level in a group. Most of the proposed feedback mechanisms measure the similarity between individuals' preference and the collective preference as her consensus level and determine whether the consensus level is acceptable [10, 12, 13]. If the group is not of acceptable consensus, then it will identify individuals who are the most incompatible with the group collective opinion, and recommend that they update their preference according to the proximity with the collective assessment [14]. Without aggregating the individual preferences into a collective preference in each round of the consensus process, Dong.et.al. proposed a peer to peer consensus reaching model which measure the consensus by using the preference proximity among individuals [15]. This kind of peer to peer consensus model can reduce the computing costs in consensus models. Additionally, when considering the cost of persuading experts to adjust opinions, some researchers proposed consensus models which provide a minimum cost advice to individuals by obtaining optimal convergence point [16, 17].

All the aforementioned decision science models are focused on the details of complex interaction process such as preference format, feedback mechanism, weight allocating, etc. In other words, they are aiming at achieving a certain level of consensus in a group from a microcosmic perspective. However, the opinion polarization of a group is a common phenomenon under a scenario with open communication, competition or antagonism [18-20]. On the one hand, it is well known that individuals who hold incompatible opinions may hardly revise themselves towards the opponents. But on the other hand, some differences between experts can still be reduced through effective communication and opinion clusters will be generated spontaneously at the stable state. In this paper, consensus which can be improved by narrowing the reducible differences is called **potential consensus**. Generally, when solving GDM problems, though total consensus may not be achieved, the potential consensus can be improved and the stability of the opinions can be reached. Thus, a GDM problem solving framework which can encourage decision makers to improve the potential consensus and then drive the group opinion reach a stable state is very valuable.

The opinions of a group can evolve into a stable state as a general final state and this process has been well studied by opinion dynamics researchers [21]. The prior efforts on the opinion dynamics process were always associated with complex networks, in which the social ties among individuals are modeled by means of links of nodes. According to the format of preference (opinion), they can be divided roughly into three types: 1) Binary format opinions: Ising model [22] and Sznajd model [23]; 2) Discrete format opinions: voter model [24, 25] and Majority Rule model [26, 27]; and 3) Continuous opinions: the bounded confidence model where individuals can adjust their viewpoint is restricted by bounded confidence are applied widely [28-31].

In bounded confidence model, an agent's opinion can be impacted by their neighbors whose opinion differ from her no more than a confidence level and the opinion profile may evolve into more than one cluster in the end where the agents in the same cluster hold the same opinion. Additionally, the static network topology is another feature of these studies. However, when individuals effect one another to adopt new opinions, the similarities between pairs of individuals in the group will be influenced. The structure of the relationship network will adapt to this change accordingly: new links between similar individuals form while links between dissimilar individuals decay. Meanwhile, in the context of the opinion interactions, individuals in a group may become more like-minded because they are connected in the network[29] or have no interaction chance because of their segregation. Thus, the network structure which describe the cooperative/hostile relationship can affect the opinion revise process in turn. Such a synchronous interplay of these two scenarios motivates co-evolving opinion dynamics studies. Holme and Newman considered a discrete opinion model that combine these two scenarios: 1) individual's opinion will follow one of her neighbor with probability $\phi$; 2) individual will rewire its connection to a new individual holding the same opinion as her with probability $1-\phi$, and explored the dynamic properties with varied $\phi$ [32]. Luo.et.al, established a model to investigate knowledge diffusion process among adaptive network, where an extended bounded confidence model is applied and individuals will rewire choose their new neighbors to seek better chance of exchange with a certain probability [33]. Su.et.al extended traditional HK model by designing reviewing rule and compared the evolution of opinions on adaptive networks and that on static networks [34]. Most of the previous opinion dynamics literatures mentioned above focused on discussing the issues at the macroscopic level, such as the topology features of complex networks and opinion survived thresholds in large scale networks which usually contains more than thousands of nodes. Furthermore, the interaction process among the individuals are always simplified, e.g. the updated opinion of an individual might be directly adapted from one of her/his neighbors' opinion [35] or be the average opinions value of neighbors [34].

Whereas the size of the group is usually much smaller in the management or business decision environment, there might be higher possibility for more complex interaction among individuals, which is similar to negotiations with compromise and persuasion. As mentioned before, this kind of interaction process is a core issue in GDM literatures. The opinion adjusting process can be affected by many factors, such as the weights of the individuals[36], the personality (degree of cooperation or selfish) [37] and the similarity of the opinions [38-40] et al. Meanwhile, the relationships in such small scale networks can be impressible, e.g. the relationship between two individuals may transform from hostile into supportive as their opinion distance becomes much closer after a negotiation. In addition, this kind of change can also influence the extent of opinion modification in the following interaction process. Thus, in this context, exploring the opinion dynamics process from a microcosmic perspective is of practical significance.

In this study, we explore this co-evolution opinion dynamics in a novel GDM framework considering the interactively influence of the changeable relationship and opinions. The objective of this framework is to provide a stability reaching group decision making model to improve the potential consensus among a group considering the dynamic relationship among decision makers. To guarantee the satisfaction of decision makers in the consensus reaching process, we not only use the bound of confidence to eliminate individuals whose opinions are too far away from each other to interact, but also set a bound of consensus to reserve the negligible disagreement among the like-minded individuals. By this way, this GDM framework can seek common ground while reserving differences on the opinions of the group. In the opinion update process, we take the personality of individuals as well as their weight as two influence factors into consideration. Furthermore, the weight of individuals which associated to the network topology will evolve after each round of interaction in this framework. In addition, we test the features of this framework through simulation experiments where the results provide deeper insights for the co-evolving dynamic model.

The paper is organized as follows. Section 2 introduces some preliminaries about GDM problems, complex networks and bounded confidence models. Section 3 represents the stability reaching GDM framework. Section 4 shows the results of simulations. The conclusions are provided in Section 5.

# 2 Preliminaries

The GDM problem is simplified to an evaluation problem where experts give one-dimensional opinions. Let $N = \{1, 2, \cdots, n\}$ and $M = \{1, 2, \cdots, m\}$ denote elements in two sets. Considering $n$ individuals (decision makers) $E=\{e_1, e_2, \cdots, e_n\}$ taking part in a GDM process with discrete time $t \in \mathbb{N}$. The individual opinion of $e_i$ at time $t \in \mathbb{N}$ is labeled as $o_i^t \in [0,1]$, and $O^t$ is called *opinion profile* at time $t \in \mathbb{N}$.

## 2.1 The Network model

A network is an abstract representation of a set of individuals and the relationships among them. Here, the set of individuals is $E$ and the relationships are described by adjacency matrix $A^t = \left(a_{ij}^t\right) \in \{0,1\}^{n \times n}$, $(i, j \in N)$, where $a_{ij}^t = 1$ when link between $e_i$ and $e_j$ exists at time $t$, and $a_{ij}^t = 0$ otherwise. Let $K^t = (k_1^t, k_2^t, \cdots, k_N^t)^{\mathrm{T}}$ be the nodes' (individuals') degree at time $t$, where $k_i^t = \sum_{j=1, j \neq i}^{n} a_{ij}^t$ is the number of neighbors of individual $e_i$, $k_i^t \in [0, n-1]$. To represent the relationships among individuals in real situations, we consider three kinds of complex networks, complete networks, scale-free networks and community networks to describe manifold initial relationships respectively.

### 1) Complete networks

When decision makers share a similar background, they might be familiar with each other and may have roughly the same authority. Then the relationships between them are represented as a complete network where every pair of individuals is joined by a link. To represent large GDM problems, we set the group size from 10 to 100 divided by 10. In this condition, the initial opinion profile $O^0$ obeys a random uniform distribution.

### 2) Scale-free networks

In large GDM problems, a common circumstance is that there are some senior or superior individuals who should have higher initial authority and weight. We denote this by the scale-free network which is characterized by the power-law degree distribution $p(k) \sim k^{-\gamma}$ with $\gamma \in [2,3]$ [41]. In scale-free networks, a minority of the nodes has large degree and the majority have smaller degree. This important property of scale-free network models most of the real networks very well.

The scale-free network generating algorithm is as follows. Consider $m_0$ vertices at the beginning (they are connected randomly), when each new node introduced into the network, $m$ links will be built $m \leq m_0$ with those old nodes taking the preferential attachment strategy. The possibility of old nodes $e_i$ being connected is proportional to its degree $k_i$: $\Pi_i = k_i / \sum_j k_j$, where $\sum_j k_j$ represents the accumulated degrees of all nodes in the network. We set $m = m_0 = 4$ here. In this case, the initial opinion profile $O^0$ also follows a uniform random distribution.

### 3) Community Network

In GDM problems, when individuals or organizations come from different academic fields or departments, the opinions can be obvious diverse. A network with a community structure can fit this kind of situations. In such networks, nodes are more likely to be connected to each other within a community than they are to nodes outside of the community [42, 43]. The network generating algorithm assumes the following: the network is composed of several communities which are started as fully connected. According to the definition of community network, we set the probability of an internal connection to 0.9, and the probability of a connection with the other community is 0.1. Namely, the link within a community will be rewired to nodes in other communities with probability 0.1. At the end of link rewiring process, the network will form a community network structure.

Without the loss of generality, we assume that the group of decision makers come from two different communities, and the number of nodes in each community is $n/2$.

The initial opinion profile $O^0$ distribution obeys a normal distribution within the community with mean parameters 0.25 and 0.75 respectively, the standard deviation parameter is 0.1 for both of them. Thus we artificially divide individuals into two factions.

## 2.2 Bounded confidence model

In general, it will be difficult for an individual to absorb others' incompatible judgements. Hence, the bounded confidence model is widely applied in opinion dynamics studies. The DW model [44] and HK model [28] are two representative continuous opinion dynamics models. In the DW model, if the opinion distance between $e_i$ and $e_j$ who are chosen randomly is less than the bound of confidence $\varepsilon$, then they will communicate and update their opinions according to

$$\begin{cases} o_i^{t+1} = \alpha \cdot o_i^t + (1-\alpha) \cdot o_j^t \\ o_j^{t+1} = \alpha \cdot o_j^t + (1-\alpha) \cdot o_i^t \end{cases} \quad (1)$$

where the positive constant $\alpha$ is a predefined convergence parameter. The confidence threshold reflects an individual's tolerance toward the alien view. Within contrast to the opinion updating rule of DW model, in HK model, individual $e_i$ will updates $o_i^t$ by averaging all opinions of her/his neighbors whose opinion lie in the confidence range of $e_i$. Specially, in a traditional HK model, all individuals have an identical confidence threshold. In these two models, the opinion profile $O^t$ will converge into clusters in the dynamic process, the cluster is where the individuals hold the same opinion.

However, in a GDM problem, we might not need the opinions converge to the same value but within a small range which means the opinions are close enough [2]. This kind of soft consensus is easier to reach and more practical. Meanwhile, the opinion dynamics of a GDM problem have a much smaller group size and a more complicated negotiation process. The convergence parameter $\alpha$ could be changeable over time and allowed to vary among the individuals. In the next section, we will introduce the opinion dynamic mechanism.

# 3 Stability reaching group decision making framework

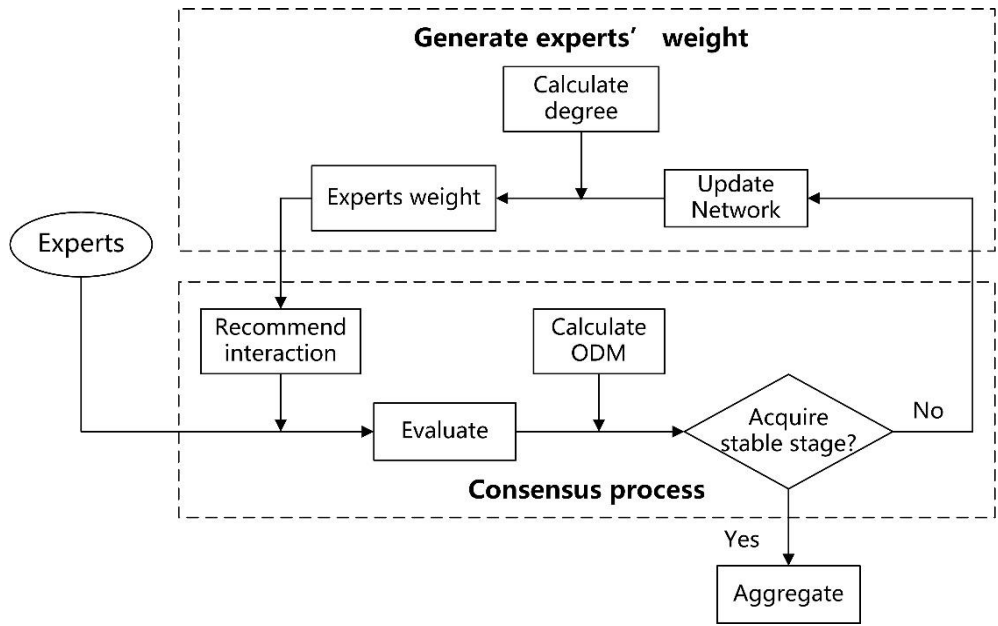

Fig.1. Stability reaching group decision making framework

In the GDM process that presented in **Fig.1**, let $o^0$ be the initial judgments that experts give. Then we use the opinion distance matrix $D^t = \left(d_{i,j}^t\right)_{n \times n}$ to represent the proximity of opinions between individuals, where

$$d_{i,j}^t = |o_i^t - o_j^t| \tag{2}$$

is the distance between $e_i$ and $e_j$ at time $t$. On the one hand, individuals prefer to take opinions of others into account if they are closer than a bound of confidence to their own opinion. Similar with the traditional DW model, the bound of confidence in our model is also determined by the tolerance threshold $\varepsilon (\varepsilon \in [0,1])$ [44]. Recommendations only happens when $d_{i,j}^t \leq \varepsilon$, while $d_{i,j}^t > \varepsilon$, $e_i$ and $e_j$ will not have interaction in terms of their incompatible judgements. On the other hand, as we have discussed, different with conventional opinion dynamic models, in the context of GDM problems, the state of agreement in a group always require individuals exhibit a state of common feeling as to the opinions towards a question. This kind of common feeling do not require opinions converge to a certain value strictly but to a bound, when the opinion distance of a pair of individuals within this bound, the opinions they hold are regarded as *compatible* and there is no necessary for them to compromise. Thus, in addition to the bound of confidence $\varepsilon$, we have the bound of consensus $\varphi$ to control the close level of opinions. When $d_{i,j}^t$ is less than threshold $\varphi$, this pairs of individuals will not be recommended changing their opinion with each other.

We donate $B\left(o_i^t, \delta\right) = \left\{o_j^t \mid d_{i,j}^t < \delta, o_j^t \in o^t\right\}$, based on the statements above, individual $e_i$ has the following **Negotiable Opinion Set**: $S_i^t = B\left(o_i^t, \varepsilon\right) - B\left(o_i^t, \varphi\right)$. The negotiable opinion set of individual $e_i$ at time $t$ contains all other opinions lie in the $[\varphi, \varepsilon)$ interval around $o_i^t$. There will be common ground for individuals $e_i$ and $e_j (o_j^t \in S_i^t)$ to compromise under this negotiable requirement. In addition, if $o_j^t \in S_i^t$, then $o_i^t \in S_j^t$, thus this kind of **Negotiable** relationship among individuals satisfies symmetric.

When $d_{i,j}^t < \varphi \vee d_{i,j}^t \geq \varepsilon, \forall i, j \in N$, interaction among individuals will be regarded as either unnecessary or impossible and the opinion dynamic process will come to end at a stable state. Thus far, we have:

**Definition 1**: If $\forall i \in N, S_i^t = \varnothing$, then opinion profile $O^t$ reaches **Stable State**.

With the confidence threshold $\varepsilon$, we are able to avoid forcing individuals who hold incompatible opinions to adjust their view and improve the potential consensus level of the group. Concretely, the bound of confidence guarantees the willingness of individuals by limiting the individuals' tolerance degree towards other opinions. Meanwhile, the bound of consensus $\varphi$ controls the consensus level to reserve the negligible disagreement among the like-minded individuals and serves as a stop condition to prevent the process from entering an infinite loop.

From **Fig.1**, we can get that if the group have not reach the **Stable State**, the network topology will evolve based on the opinion distance matrix and the individuals' weight will be changed correspondingly. Then individuals will be advised to communicate in pairs and their opinions will be updated. The new opinion distance matrix will be calculated and judged again. This process will iterate until the stable state is reached. As is well known, cooperative and hostile behaviors of individuals in a group may lead to a polarized group. In other words, the opinions of a group may converge into more than one clusters in the stable state, let $E_1^t, E_2^t, \ldots, E_m^t$ ($m \geq 1$) be the partition of set $E$ in round $t$, $E_u^t \neq \varnothing$ and $\bigcup_{u=1}^m E_u^t = E$. Then we have:

**Definition 2**: For the set of opinions in stable state, in round $T$, $O^T = \{o_1^T, o_2^T, \ldots, o_n^T\}$, there are $m$ clusters with partitions $E_1^T, E_2^T, \ldots, E_m^T$, if $\forall u \in M, \bigcap_{i \in E_u^T} B(o_i^T, \varphi) = \bigcup_{j \in E_u^T} B(o_j^T, \varphi)$.

In traditional bounded confidence models, the individuals in one cluster must reach complete consensus (if and only if they hold exactly the same opinion) [28], which is a strict condition and is difficult to achieve in daily decision making process. However, as soft consensus is applied in the proposed model, individuals are in one cluster only if the opinion distance close enough (not exceed an threshold $\varphi$). It's easy to see that the condition of cluster in the traditional bounded confidence model is a special case ($\varphi = 0$) of the proposed model.

Since there could be more than one opinion cluster, there is an aggregation process to derive the collective judgement in the end. That is the whole process of the stability reaching group decision making framework.

## 3.1 Evolution of network structure

It is reasonable to assume that an individual will be more convincing if her/his opinion is supported by more people. Following this idea, we adjust the relationship network of individuals according to the opinion distances: more similar ideas in the network can lead to more links which can determine a higher weight for a certain individual. Thus, the individuals' weights can be evaluated objectively through dynamic graph topology.

### 1) Network rewiring rule

The links between individuals is updated according to their current opinion distance condition. If the opinion distance $d_{i,j}^t \geq \varepsilon$, the link between $e_i$ and $e_j$ will be removed which means and these individuals who hold incompatible opinions do not support but hostile to each other. If $d_{i,j}^t < \varphi$, then $e_i$ and $e_j$ will be connected and build a cooperate relationship through interactions. Else when $d_{i,j}^t \in [\varphi, \varepsilon)$, the relationship between individuals do not have obvious change and will remain the same. As can be inferred that, after a round of negotiation, the opinions of the selected pair of individuals and their opinion distance from the rest opinions might be changed. Then the links of the network will be adjusted according to the network rewiring rule. Thus the evolution of the global network topology can be driven by opinion interaction.

### 2) Weight generating process

$k_i^t$ which represents the degree of node $e_i$ will be greater than $k_j^t$ when there are more individuals support $o_i^t$. Consequently, the degree $k_i^t$ of $e_i$ means the relative authority of an individual. We regard $K^t$ as the decision makers' weights at step $t$. In this framework, the initial relationship as well as the dynamic authority are taken into consideration when generating the weight. The weight calculation is simple and efficient; and the disadvantage of static weights where in the initial authority manipulates the result can be avoided also.

## 3.2 Evolution of pairwise opinions

This part mainly introduces the process of the recommendation strategy and opinion updating rule.

### 1) Recommendation strategy

In the framework, $e_j$ will be recommended to negotiate with $e_i$ when their opinion distance satisfies $d_{i,j}^t = \max\limits_{s \in N, s \neq i} \left\{ d_{i,s}^t \mid d_{i,s}^t \in [\varphi, \varepsilon] \right\}$. In other words, the most conflict pair of opinions which satisfies the negotiable requirement will be selected. This strategy ensures the efficiency of the opinion revising process. Specially, if there are more than one pair of individuals share the same opinion distance as $\max d_{i,s}^t \in [\varphi, \varepsilon)$, they will be chosen randomly to have communication.

**Remark 1**: It is easy to know that the more incompatible opinions have the greater potential space for consensus improvement [45] but may be the least likely to agree. In this model, the advantage of our recommendation strategy is that it distinguishes the individuals who share the greatest opinion distance and also considers the individual's tolerance at the same time. Thus, the efficiency of opinions' stable state reaching can be increased which is beneficial for converging to cluster(s) and reaching stable state, and the satisfactory of individual can be guaranteed in this model.

### 2) The rule of opinion updating

The relative weight $k_i^t$ can affect the compromise level of opinions in a negotiation, individuals with large weight can be more persuasive. Except for $k_i^t$, individual's persistency degree $p$ can also influence the revise process. The greater $p$ is, the more persistent or unbending the individuals will be. It means they will keep more individual opinion and will only take a small step closer to the other. Whereas, they tend to be swayed by others.

Combined with these two factors, the algorithm of generating the individual change of view is an extension of revise/advise model in [15]:

$$\begin{cases} \alpha_i^t = 1 - \dfrac{k_j^t}{(k_i^t + k_j^t) \cdot p} \\ \alpha_j^t = 1 - \dfrac{k_i^t}{(k_i^t + k_j^t) \cdot p} \end{cases}, \qquad (3)$$

the value of $\alpha_i^t$ changes with respect to $k_i^t$ and $k_j^t$. Particularly, when $k_i^t = k_j^t = 0$, we regard $k_i^t = k_j^t = 1$ based on the initial significance of relative weight, then $\alpha_i^t = \alpha_j^t = 1 - 1/(2p)$. The Eq.(3) has some important properties worth noting: 1) $\alpha_i^t \in [1 - 1/p, 1)$, the parameter $p$ in our model is assigned by 2, 3 and 4, thus the change of opinion will not be too much which can guarantee the satisfaction of individuals. 2) when $k_i^t = k_j^t$, the opinion revise level is $1 - 1/(2p)$ for $e_i$ and $e_j$, they will have equal transformation at this time. 3) if the relative weight of $e_i$ greater than $e_j$, the former will have less change than the later which corresponds to common sense, convergence parameter $\alpha_i^t$ can describe the present persuade power of individual $e_i$ accordingly.

The opinions in next round will be the following:

$$\begin{cases} o_i^{t+1} = \alpha_i^t \cdot o_i^t + (1 - \alpha_i^t) \cdot o_j^t \\ o_j^{t+1} = \alpha_j^t \cdot o_j^t + (1 - \alpha_j^t) \cdot o_i^t \end{cases}. \qquad (4)$$

Besides, from the Eq.(3) and (4), we can easily have two inequalities that 1) the order of these two interacting opinions is preserved, that is $o_i^{t+1} - o_j^{t+1} > 0$ if $o_i^t - o_j^t > 0$, 2) $d_{i,j}^{t+1} < d_{i,j}^t$. These imply an important property of the proposed interaction rule that individuals' opinions will become closer to each other but never intersect. It is similar to the conferring process in real word GDM problems.

**Remark 2:** In this model, as have mentioned, the opinion distance affects the structure of the relationship network by influencing the rewire process with parameters $\varepsilon$ and $\varphi$. Interactively, the relationship network can also be a significant factor to the opinion adjusting process through the dynamic weight generated from the objective network structure. By this mechanism, the dynamic relationship can be described and its influence to the opinion adjusting process is also reflected.

*Theorem 1:* In the proposed stability reaching model, the opinion distance between interacting individuals can be effectively reduced by $1 - 1/p$ in each step, which is inversely proportional to individual's own weight. Especially, if one of the individual's weight equals to zero, only this individual's opinion will be revised.

From *Theorem 1*, we have that the adjust level depends on individual's degree of stubbornness $p$, the convergence speed will slow down with a higher $p$.

*Theorem 2:* In this framework, the stable state will be reached in a limited number of steps.

There are two significant factors that guarantee the amount of time steps is finite. Firstly, the recommendation strategy is picking up the two individuals whose opinion distance satisfies $\max_{i,j \in E}\{d_{i,j}^t\} \in [\varphi, \varepsilon)$. This strategy contributes to the most efficient opinion adjusting. Secondly, threshold $\varphi$ limits the required consensus level, the iteration will stop once the end condition be reached.

## 3.3 Aggregation

Considering there may be more than one cluster at the stable state, we choose the weighted arithmetic means to aggregate opinions:

$$O = \frac{O^T \cdot K^T}{\sum K^T}. \qquad (5)$$

After this step, the opinions in the stable state are aggregated to a certain value and thus the final solution of GDM problem can be generated.

# 4 Simulation Analysis

In this section, we present and analyze our simulation results. First, we show the necessary time steps of reaching the stable state under different parameter $p$ in various initial network (Complete network, Scale-free network and Community network). Then investigate the number of the opinion clusters in the stable state with different parameters' value. All the simulation experiments are implemented by mathematical software MATLAB R2013a.

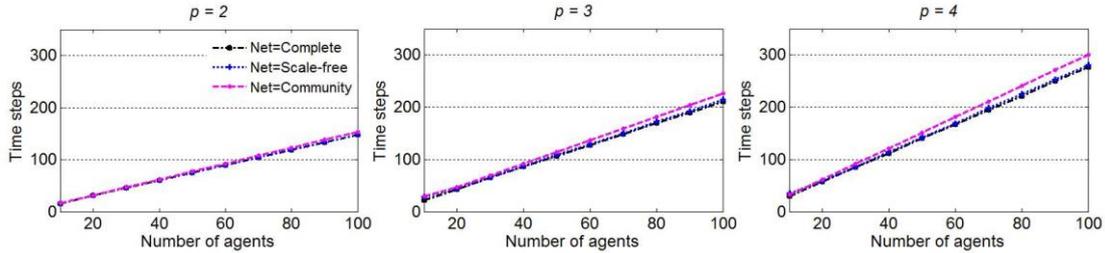

Fig.2 Number of time steps with $p = 2, 3, 4$ in different initial network ($\varepsilon = 0.5$, $\varphi = 0.1$, iteration = 100).

**Fig.2** shows the impact of individual's persistency $p$ and the number of individuals with different initial relationship on time steps. The group size $n$ is set from 10 to 100 with an increase of 10 individuals each time. The bound of confidence $\varepsilon$ is set to 0.5 and the required consensus level $\varphi$ is 0.1. We ran the simulation experiments in the three kinds of networks with 100 iterations for each group of parameters.

From **Fig.2**, we can observe the time cost with $p = 2, 3, 4$ respectively. Basically,

the graph shows a linear relationship between total steps and the size of groups. More specifically, the time steps ascend as the number of individual increases. The total steps will be more with greater $p$. From Eq.错误!未找到引用源。, we have $d_{i,j}^t = \max\limits_{s \in N, s \neq i} \{d_{i,s}^t | d_{i,s}^t \in [\varphi, \varepsilon]\}$ will be reduced by $1-1/p$ in time $t+1$, thus this parameter influences the total time steps by deciding the opinion change degree. As can be identified in **Fig.2**, the time cost seems relevant to the initial network class. The black line and the blue line, which represent the complete network and scale-free network respectively, coincide with each other well. While the red line embodies community network is always above the former and the gap between them becomes more distinct as $p$ grows. The reason for this phenomenon is that opinions are assigned by the community partition in the community network and there are fewer connections between individuals in different communities, these result in a kind of barrier for prospective interaction which contributes to a longer process before forming a complete network and reaching consensus compared with the other networks. Thus, when the opinion convergence speed slows down, the community network will be effected the most. Owing to this fact, there will be an adjustment in adjacency matrix according to the opinion distance after the opinions are assigned, the difference on network structure is weakened. Then it can explain why the blue line and black lines, represents complete network and scale-free network respectively, coincide well in **Fig.2**.

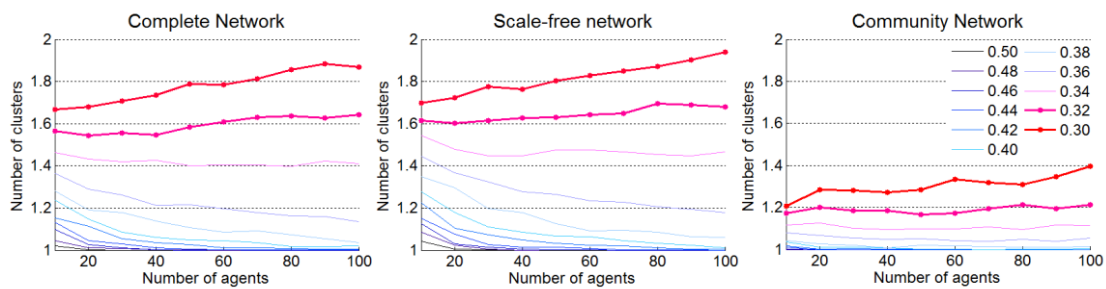

Fig.3 The effect of bounded confidence $\varepsilon = 0.3$, $^{+0.02}_{...}$, $0.5$ on the number of clusters with different $n$ ($p = 2$, $\varphi = 0.1$, iteration = 1500).

Then we explore the impact of bounded confidence threshold $\varepsilon$ on clusters' amount $m$ in the framework proposed. The bound of confidence is set from 0.3 to 0.5 with 0.02 increase each time. The consensus level $\varphi$ is assigned to a constant value 0.1. We applied the simulation experiments in complete network, scale-free network and community network respectively. **Fig.3** shows the result and every value in it is the average value of the result of 1500 simulations to reduce stochastic error.

The result is similar to Deffuant [44], Hegselmann [28] and Alizadeh [46] reached in their study: the high confidence thresholds lead opinions towards an global consensus in the stable state, but low thresholds result in several opinion clusters. As it can be observed in **Fig.3**, the warm color lines which represent a relatively low $\varepsilon$ are always above those cold color lines which describe high bound of confidence. And there is no intersection between those lines in any kind of network. This phenomenon indicates that it will be easier to format global consensus when individuals are more tolerant with others view and willing to have interaction. Otherwise, more opinion clusters will come into being in the end.

From the trend of lines in **Fig.3**, we can note that $m$ is descending along with the growth of group size. The result is counter-intuitively that more initial opinions existing in a network contributes to global consensus reaching on the contrary. It is because there will be more coordinators or bridge nodes who can guide extreme opinions to converge. It can explain the phenomenon for those cold color lines drop slightly with the increasing group size. Interestingly, however, not all the lines obey this rule, the two bold lines which display the results when $\varepsilon$ is 0.3 and 0.32 respectively climb up instead. That is because the individuals' tolerance of other views become smaller, and this results in the fact that the region of interactive opinions narrowed, then more individuals means more rejected opinions in such condition. Meanwhile, the number of coordinators will be reduced spontaneously and then the group will split into more opinion clusters in the stable state. Let us assume an extreme situation that $\varepsilon = 0$, then $m$ will be as much as the number of different initial opinions. Thus, it can be inferred that the slope of lines will be higher with less $\varepsilon$.

The results also have some contradictions in various networks. The number of clusters in the community network is significantly lower compared with the first two graphs. And the lines in the second graph of **Fig.3** are slightly higher than those in the first. The cause of the former result is the special net structure and the initial opinion distribution in community network. Since the individuals' weight in a community are similar and the opinion distances are relatively smaller, those inner opinions have hardly obstructions to converge. Meanwhile, the rule of opinion dynamics results in more coordinators with higher weight existing in communities who can drive a global consensus reaching. In addition, there is no opinion gap between the communities in our model, which makes the opinion adjusting between communities much easier. These conditions all drive the global consensus reaching to become more likely. However, in real cases, conditions would be much different. A gap will always exist between communities which makes the consensus reaching more difficult. Also, the size of communities can vary greatly which leads to very different nodes' degree distribution, the lower weight of coordinators can make them less convictive and lead to more clusters in the end. Thus from the simulation results and the analysis, we can draw the conclusion that the global consensus can be reached easier when the community has a similar scale and has no opinion gap. In terms of the later phenomenon that cluster amount in a complete network is less than that in a scale-free network, it is due to the individuals' initial authorities have less difference in the former than in the latter. As can be seen from Eq. (3), the equal weight of individuals (degree of nodes) in a complete network means smaller gap of persuade power ($\alpha^t$), then individuals can play the coordinator role better which is contrary to scale-free network. Thus the cluster amount $m$ in complete network is slightly less than that in scale-free network.

To conclude, the number of clusters will become greater with a less confidence threshold $\varepsilon$. Besides, when $\varepsilon \geq 0.34$, larger group size $n$ can lead to less clusters, but when $\varepsilon < 0.34$, there will be more clusters come into being with greater $n$. Finally, opinions tend to converge under full and free discussion, it will be effective to reach a global consensus if those coordinate individuals have high authority.

# 5 Conclusions

This study has explored the opinion dynamic mechanism in a GDM framework where the relationship of individuals is changeable depending on the initial relationship network as well as the current opinion distances. It has shown that the opinions will evolve into more than one cluster through improving the potential consensus. Further, through simulation experiments and analysis, we show that the converge time and the cluster amount in the final stage can be influenced by the initial structure of the network and the opinion distribution, which attests the significance of this novel framework based on a co-evolving network. Specifically, the findings can be summarized as follows: 1) The low persistence degree of individuals can reduce the time steps of negotiation; 2) Greater bound of confidence $\varepsilon$ contributes to global consensus making; 3) A larger group size $n$ can lead to less cluster amount $m$ in the stable state when $\varepsilon \geq 0.34$, but lead to more clusters when $\varepsilon < 0.34$. The results of this study indicate that to lead a global consensus to the final stage, the organizers are supposed to reduce the initial weight difference and raise the individuals' confidence threshold; recognize and set more weight on bridge nodes when extreme opinion exists or more clusters can be predicted.

This work enriches the existing GDM system by utilizing the complex networks to represent the changeable relationship which can help to evaluate the dynamic weight objectively and exploring some interesting behavior in such co-evolution framework under the stability guiding process. A further study investigating the opinion dynamics mechanism of multiple attribute GDM problems where the relationship network can be described by multilayer networks would be very interesting.

[2] Herrera-Viedma E, Cabrerizo FJ, Kacprzyk J, Pedrycz W. A review of soft consensus models in a fuzzy environment. Information Fusion. 2014;17:4-13.

[3] Surowiecki J. The wisdom of crowds: Anchor; 2005.

[4] Carlsson C, Ehrenberg D, Eklund P, Fedrizzi M, Gustafsson P, Lindholm P, et al. IFORS-SPC Conference on Decision Support SystemsConsensus in distributed soft environments. European Journal of Operational Research. 1992;61:165-85.

[5] Kim SH, Choi SH, Kim JK. An interactive procedure for multiple attribute group decision making with incomplete information: Range-based approach. European Journal of Operational Research. 1999;118:139-52.

[6] Wu Z, Xu J. A consistency and consensus based decision support model for group decision making with multiplicative preference relations. Decision Support Systems. 2012;52:757-67.

[7] Herrera-Viedma E, Alonso S, Chiclana F, Herrera F. A consensus model for group decision making with incomplete fuzzy preference relations. IEEE Transactions on fuzzy Systems. 2007;15:863-77.

[8] Dong Y, Xu Y, Li H, Feng B. The OWA-based consensus operator under linguistic representation models using position indexes. European Journal of Operational Research. 2010;203:455-63.

[9] Herrera-Viedma E, Martinez L, Mata F, Chiclana F. A Consensus Support System Model for Group Decision-Making Problems With Multigranular Linguistic Preference Relations. IEEE Transactions on Fuzzy Systems. 2005;13:644-58.

[10] Wu J, Chiclana F, Herrera-Viedma E. Trust based consensus model for social network in an incomplete linguistic information context. Applied Soft Computing. 2015;35:827-39.

[11] Altuzarra A, Moreno-Jiménez JM, Salvador M. Consensus building in AHP-group decision making: a Bayesian approach. Operations research. 2010;58:1755-73.

[12] Herrera-Viedma E, Herrera F, Chiclana F. A consensus model for multiperson decision making with different preference structures. IEEE Transactions on Systems, Man, and Cybernetics-Part A: Systems and Humans. 2002;32:394-402.

[13] Mata F, Martínez L, Herrera-Viedma E. An adaptive consensus support model for group decision-making problems in a multigranular fuzzy linguistic context. IEEE Transactions on fuzzy Systems. 2009;17:279-90.

[14] Cabrerizo FJ, Alonso S, Herrera-Viedma E. A consensus model for group decision making problems with unbalanced fuzzy linguistic information. International Journal of Information Technology & Decision Making. 2009;8:109-31.

[15] Dong Q, Cooper O. A peer-to-peer dynamic adaptive consensus reaching model for the group AHP decision making. European Journal of Operational Research. 2016;250:521-30.

[16] Ben-Arieh D, Easton T. Multi-criteria group consensus under linear cost opinion elasticity. Decision Support Systems. 2007;43:713-21.

[17] Gong Z, Zhang H, Forrest J, Li L, Xu X. Two consensus models based on the minimum cost and maximum return regarding either all individuals or one individual. European Journal of Operational Research. 2015;240:183-92.

[18] Proskurnikov A, Matveev A, Cao M. Consensus and polarization in Altafini's model with